\documentclass[onecollarge,natbib]{svjour2}
\bibpunct{[}{]}{;}{n}{}{,} % to get "[numbered]" references from natbib
\smartqed  % flush right qed marks, e.g. at end of proof
\usepackage{graphicx}
\usepackage{epsfig}
\newcommand\ba{\begin{eqnarray}}
\newcommand\ea{\end{eqnarray}}

\newcommand{\be}{\begin{equation}}
\newcommand{\ee}{\end{equation}}

\newcommand{\bas}{\begin{eqnarray*}}
\newcommand{\eas}{\end{eqnarray*}}

\newcommand{\bno}{\begin{eqnarray*}}
\newcommand{\eno}{\end{eqnarray*}}

\journalname{Few-Body Systems}

\begin{document}

\title{In-Medium Pion Valence Distribution Amplitude
\thanks{This work was partly supported by the Funda\c c\~ao de Amparo \`a Pesquisa do Estado de
S\~ao Paulo (FAPESP) and Conselho Nacional de Desenvolvimento 
Cient\'ifico e Tecnol\'ogico  (CNPq) of Brazil.}
}



\author{K.~Tsushima         \and
        J.~P.~B.~C~de~Melo %etc.
}


\institute{
K. Tsushima \at
Laborat\'orio de F\'\i sica Te\'orica e Computacional, 
Universidade Cruzeiro do Sul, 01506-000 S\~ao Paulo, Brazil \\
              \email{kazuo.tsushima@gmail.com, kazuo.tsushima@cruzeirodosu.edu.br}
              %Tel: +55-11-33853004
              %\email{fauthor@example.com}           %  \\
           %\and
            %\at
\vspace{1ex}
\\
J.~P.~B.~C~de~Melo \at
Laborat\'orio de F\'\i sica Te\'orica e Computacional, 
Universidade Cruzeiro do Sul, 01506-000 S\~ao Paulo, Brazil \\            
              \email{joao.mello@cruzeirodosul.edu.br}
}

\date{Received: date / Accepted: date}

\maketitle

\begin{abstract}
After a brief review of the quark-based model for nuclear matter,
and some pion properties in medium presented in our previous works, 
we report new results for the pion valence wave function as well as 
the valence distribution amplitude in medium, which are presented in our recent article. 
We find that both the in-medium pion valence distribution            
and the in-medium pion valence wave function, are substantially  
modified at normal nuclear matter density, due to the reduction in the pion decay constant.
\end{abstract}

\section{Introduction}
\label{intro}
%{\it Introduction:}~ 
One of the most exciting and challenging topics in hadronic physics is 
to study the modifications of hadron properties in nuclear medium, 
and to study how such modifications give impact on observables in medium.
Since hadrons are composed of quarks, antiquarks and gluons, one can naturally expect that 
hadron internal structure would change when they are immersed in nuclear 
medium~\cite{Brown,Hatsuda,Saito2007,Hayano,Brooks}.
This question, becomes particularly interesting when it comes to that of pion. 
To be able to answer it, one first needs, simpler, 
effective quark-antiquark models of pion, which are successful in describing  
its properties in vacuum. Among such models, light-front constituent quark model 
has been very successful in describing the electromagnetic form factors, electromagnetic radii 
and decay constants of pion and 
kaon~\cite{deMelo1997,deMelo1999,deMelo2002,daSilva2012,Yabusaki2015,deMelo:2015yxk}.
Recent advances in experiments, indeed suggest to make it possible 
to access to the pion properties in a nuclear 
medium~\cite{Saito2007,Hayano,Brooks,deMelo2014,pimedium2}.

Among the all hadrons, pion is the lightest, and it is believed as 
a Nambu-Goldstone boson, which is realized in nature emerged by  
the spontaneous breaking of chiral symmetry. 
This Nambu-Goldstone boson, pion, plays very important  
and special roles in hadronic and nuclear 
physics.
However, because of its special properties, particularly the unusually light mass,  
it is not easy to describe the pion properties in medium as well as in vacuum  
based on naive quark models, even though such models can be successful in describing 
the other hadrons. 

Recently, we studied the properties of pion in 
nuclear medium~\cite{deMelo2014,pimedium2,piDA}, 
namely, the electromagnetic form factor, charge radius and weak decay constant, 
distribution amplitude (DA) by using a light-front constituent quark model. 
There, the in-medium input was calculated by the quark-meson coupling (QMC) 
model~\cite{Saito2007,Guichon}.
The main purpose of this article is, to report our new results for 
the in-medium pion DA studied in Ref.~\cite{piDA}, supplemented by the other 
pion properties already presented in Refs.~\cite{deMelo2014,pimedium2}.

%{\it The QMC Model}:~
Below, we first briefly review the QMC model, 
the quark-based model of nuclear matter, to study the pion properties in medium.
The QMC model was invented by Guichon~\cite{Guichon} 
to describe the nuclear matter based on the quark degrees 
of freedom. The self-consistent exchange of the scalar-isoscalar $\sigma$ and 
vector-isoscalar $\omega$ mean fields coupled directly to 
the relativistic confined quarks, are the key and novelty for the new  
saturation mechanism of nuclear matter. The model was extended to finite nuclei~\cite{QMCfinite}, 
and has successfully been applied for various nuclear and hadronic phenomena~\cite{Saito2007}.

The effective Lagrangian density for a uniform, spin-saturated,
and isospin-symmetric nuclear system (symmetric nuclear matter)
at the hadronic level is given by~\cite{Saito2007,Guichon}, 
\begin{equation}
{\cal L} = {\bar \psi} [i\gamma 
\cdot \partial -m_N^*({\hat \sigma}) -g_\omega {\hat \omega}^\mu \gamma_\mu ] \psi
+ {\cal L}_\textrm{meson} ~,
\label{lag1}
\end{equation}
where $\psi$, ${\hat \sigma}$ and ${\hat \omega}$ are respectively the nucleon,
Lorentz-scalar-isoscalar $\sigma$, and Lorentz-vector-isoscalar $\omega$ field operators with, 
\begin{equation}
m_N^*({\hat \sigma}) \equiv m_N - g_\sigma({\hat \sigma}) {\hat \sigma}.
\label{effnmass}
\end{equation}
Note that, in symmetric nuclear matter isospin-dependent $\rho$-meson mean filed 
is zero, and thus we have omitted it.
Then the relevant free meson Lagrangian density is given by,
\begin{equation}
{\cal L}_\mathrm{meson} = \frac{1}{2} (\partial_\mu {\hat \sigma} 
\partial^\mu {\hat \sigma} - m_\sigma^2 {\hat \sigma}^2)
- \frac{1}{2} \partial_\mu {\hat \omega}_\nu (\partial^\mu 
{\hat \omega}^\nu - \partial^\nu {\hat \omega}^\mu)
+ \frac{1}{2} m_\omega^2 {\hat \omega}^\mu {\hat \omega}_\mu. 
\label{mlag1}
\end{equation}
Hereafter, we consider in the rest frame of symmetric nuclear matter. 
Then, within Hartree mean-field approximation, 
the nuclear (baryon) and scalar densities are respectively given by,
\begin{eqnarray}
\rho &=& \frac{4}{(2\pi)^3}\int d\vec{k}\ \theta (k_F - |\vec{k}|)
= \frac{2 k_F^3}{3\pi^2},
\hspace{3ex}
\rho_s = \frac{4}{(2\pi)^3}\int d\vec{k} \ \theta (k_F - |\vec{k}|)
\frac{m_N^*(\sigma)}{\sqrt{m_N^{* 2}(\sigma)+\vec{k}^2}},
\label{rhobs}
\end{eqnarray}
here, $m^*_N(\sigma)$ is the value (constant)  of effective nucleon mass at given density 
(see also Eq.~(\ref{effnmass})).
In the standard QMC model~\cite{Saito2007,Guichon,QMCfinite} the MIT bag model is used, 
and the Dirac equations for the light quarks inside a nucleon (bag) composing nuclear 
matter, are given by, 
\begin{eqnarray}
\left[ i \gamma \cdot \partial_x -
(m_q - V^q_\sigma)
\mp \gamma^0
\left( V^q_\omega +
\frac{1}{2} V^q_\rho
\right) \right]
\left( \begin{array}{c} \psi_u(x)   \\
\psi_{\bar{u}}(x) \\ 
\end{array} \right) &=& 0,
\label{diracu}\\
\left[ i \gamma \cdot \partial_x -
(m_q - V^q_\sigma)
\mp \gamma^0
\left( V^q_\omega -
\frac{1}{2} V^q_\rho
\right) \right]
\left( \begin{array}{c} \psi_d(x) \\   
\psi_{\bar{d}}(x) 
\end{array} \right) &=& 0~.
\label{diracd}
\end{eqnarray}
As usual, Coulomb interaction is neglected, and SU(2) symmetry is assumed, 
~$m_{u,\bar{u}}=m_{d,\bar{d}} \equiv m_{q,\bar{q}}$. 
The corresponding effective (constituent) quark masses are defined  
by, $m^*_{u,\bar{u}}=m^*_{d,\bar{d}}=m^*_{q,\bar{q}} \equiv m_{q,\bar{q}}-V^q_{\sigma}$, 
to be explained later. 

As mentioned already, in symmetric nuclear matter within Hartree approximation, 
the $\rho$-meson mean field is zero, $V^q_{\rho}=0$,~in Eq.~(\ref{diracd}),   
and we ignore it.  The constant mean-field potentials are defined as, 
$V^q_{\sigma} \equiv g^q_{\sigma} \sigma =  g^q_\sigma <\sigma>$, and, 
$V^q_{\omega} \equiv g^q_{\omega} \omega= g^q_\omega <\omega>$,~with $g^q_\sigma$, and 
$g^q_\omega$, are the corresponding quark-meson coupling constants, 
where the quantities with the brackets stand for the expectation values 
in symmetric nuclear matter~\cite{Saito2007}. 
%Since the average velocity is zero, ~$<\bar{\psi_q} \vec{\gamma} \psi_q> = 0$,
%in the nuclear matter rest frame, no spacial-dependent source for the vector-meson 
%mean fields arise, and only the terms proportional to~$\gamma^0$ are kept in Eq.~(\ref{diracd}). 
%(More details are given in Ref.~\cite{Saito2007}.)

The same meson mean fields $\sigma$ and $\omega$ for the quarks  
in Eqs.~(\ref{diracu}) and~(\ref{diracd}), satisfy self-consistently 
the following equations at the nucleon level (see Eq.~(\ref{rhobs})):
\begin{eqnarray}
{\omega}&=&\frac{g_\omega \rho}{m_\omega^2},
%\label{omgf}\\
\hspace{3ex}
{\sigma}=\frac{g_\sigma }{m_\sigma^2}C_N({\sigma})
\frac{4}{(2\pi)^3}\int d\vec{k} \ \theta (k_F - |\vec{k}|)
\frac{m_N^*(\sigma)}{\sqrt{m_N^{* 2}(\sigma)+\vec{k}^2}}
=\frac{g_\sigma }{m_\sigma^2}C_N({\sigma}) \rho_s,
\label{osigf}\\
%C_N(\sigma) &=& \frac{-1}{g_\sigma(\sigma=0)}
%\left[ \frac{\partial m^*_N(\sigma)}{\partial\sigma} \right],
%\label{CN}
\end{eqnarray}
where $C_N(\sigma) \equiv \frac{-1}{g_\sigma(\sigma=0)}
\left[ {\partial m^*_N(\sigma)}/{\partial\sigma} \right]$ 
is the constant value of the scalar density ratio~\cite{Saito2007,Guichon,QMCfinite}.
Because of the underlying quark structure of the nucleon used to calculate
$M^*_N(\sigma)$ in nuclear medium (see Eq.~(\ref{effnmass})), 
$C_N(\sigma)$ gets nonlinear $\sigma$-dependence,
whereas the usual point-like nucleon-based model yields unity, $C_N(\sigma) = 1$.
It is this $C_N(\sigma)$ or $g_\sigma (\sigma)$ that gives a novel saturation mechanism
in the QMC model, and contains the important dynamics which originates in the quark structure
of the nucleon. Without an explicit introduction of the nonlinear
couplings of the meson fields in the Lagrangian density at the nucleon and meson level,
the standard QMC model yields the nuclear incompressibility of $K \simeq 280$~MeV with 
$m_q=5$ MeV, which is in contrast to a naive version of quantum hadrodynamics (QHD)~\cite{QHD}
(the point-like nucleon model of nuclear matter),
results in the much larger value, $K \simeq 500$~MeV;
the empirically extracted value falls in the range $K = 200 - 300$ MeV.

Once the self-consistency equation for the ${\sigma}$, 
Eq.~(\ref{osigf}), has been solved, one can evaluate the total energy per nucleon:
\begin{equation}
E^\mathrm{tot}/A=\frac{4}{(2\pi)^3 \rho}\int d\vec{k} \
\theta (k_F - |\vec{k}|) \sqrt{m_N^{* 2}(\sigma)+
\vec{k}^2}+\frac{m_\sigma^2 {\sigma}^2}{2 \rho}+
\frac{g_\omega^2\rho}{2m_\omega^2} .
\label{toten}
\end{equation}
We then determine the coupling constants, $g_{\sigma}$ and $g_{\omega}$, so as
to fit the binding energy of 15.7~MeV at the saturation density $\rho_0$ = 0.15 fm$^{-3}$
($k_F^0$ = 1.305 fm$^{-1}$) for symmetric nuclear matter.

%%%
\begin{figure}[tb]
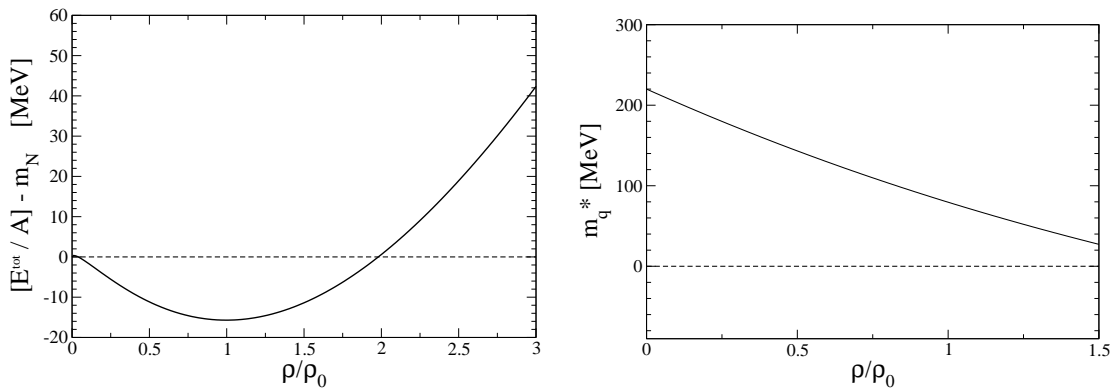

\begin{center}
%\hspace*{-1.4cm}
\mbox{
\epsfig{figure=EnergyDen.eps,width=7cm} %% ,angle=90}
\hspace{2ex}
\epsfig{figure=mqstar2.eps,width=7cm} %% ,angle=90}
} \par
\caption{Negative of the binding energy per nucleon for symmetric nuclear matter,  
$(E^{tot}/A)-m_N$, v.s. $\rho$ ($\rho_0=0.15$ fm$^{-3}$) 
with the vacuum quark mass $m_q=m_{\bar{q}}=220$~MeV ($q=u,d$), calculated by the 
QMC model~(left panel), and effective constituent quark mass~$m^*_q$ 
($q=\bar{q}=u,\bar{u},d,\bar{d}$)~(right panel). 
(Both figures are taken from Ref.~\cite{deMelo2014}.) 
The incompressibility $K$ obtained is $K=320.9$ MeV. 
\label{E/Amqstar}
}
%\vspace{1.0cm}
\end{center} 
\end{figure}
%%%%%%

In Refs.~\cite{deMelo:2015yxk,deMelo2014}, the quark mass in vacuum was used 
$m_{q,\bar{q}}~=$~220~MeV to study the pion properties in symmetric nuclear matter.
With this value the model can reproduce the electromagnetic form factor and the decay 
constant well in vacuum~\cite{deMelo2002}.
So, we use the same value in this study.
Then, all the nuclear matter saturation properties are generated by using 
this light-quark mass value.
In other words, the different values of $m_q$ in vacuum generate the corresponding 
different nuclear matter properties, except for the saturation point of the 
symmetric nuclear matter, $\rho = \rho_0$ (normal nuclear matter density, 0.15 fm$^{-3}$) with 
the empirically extracted binding energy of $15.7$ MeV.
This saturation point condition is generally used to constrain 
the models of nuclear matter.
Thus, we have obtained the necessary properties of the light-flavor constituent quarks in symmetric
nuclear matter with $m_q = 220$~MeV, namely, the density dependence of the effective mass 
(scalar potential) and vector potential.
The same in-medium constituent quark properties which reproduce the nuclear saturation properties  
will be used as input to study the pion properties in symmetric nuclear matter.

In Figs.~\ref{E/Amqstar}~we show 
negative of the binding energy per nucleon ($E^\mathrm{tot}/A - m_N$) (left-panel),
and effective constituent light-quark mass, $m_q^*$ (right panel), in symmetric nuclear 
matter~\cite{deMelo2014}.

\section{Pion properties in medium}
\label{pimedium}

%{\it The Model:}~ 
The light-front constituent quark model 
used here~\cite{deMelo2002}, although simple, 
is quite successful in describing 
the properties of pion in vacuum.

\begin{figure}[tb]
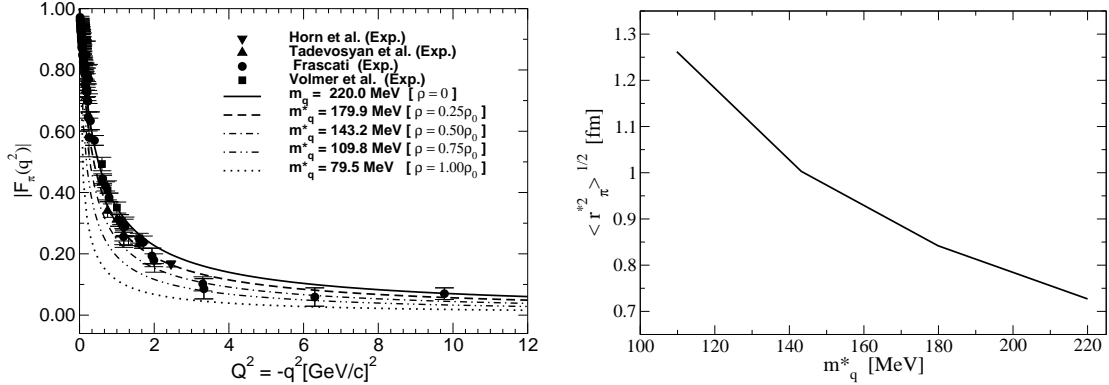

\begin{center}
%\hspace*{-1.4cm}
\mbox{
\epsfig{figure=pifofactor2.eps,width=7cm} %% ,angle=90}
\hspace{2ex}
\epsfig{figure=radius2MeV.eps,width=7cm} %% ,angle=90}
} \par
\caption{Pion charge form factor (left panel), and root mean-square 
charge radius in symmetric nuclear matter (right panel).
Both are taken from Ref.~\cite{deMelo2014}.
\label{ffradius}
}
%\vspace{1.0cm}
\end{center} 
\end{figure}

As examples, we show in Fig.~\ref{ffradius} pion charge form factor (left panel) 
and root mean-square pion charge radius in symmetric nuclear matter 
calculated in Ref.~\cite{deMelo2014}.
%%%

Here, we note that the pion mass up to normal nuclear matter density is
expected to be modified only slightly, where the modification $\delta m_\pi$ 
at nuclear density $\rho = 0.17$ fm$^{-3}$, averaged over the pion isospin states, 
is estimated as $\delta m_\pi \simeq +3$ MeV~\cite{Hayano,Kienle,Meissner,Vogl,Lutz}.
Therefore, we approximate the effective
pion mass in nuclear medium to be the same as in vacuum,
$m^*_\pi = m_\pi$, up to $\rho = \rho_0 = 0.15$ fm$^{-3}$, the maximum 
nuclear matter density treated. %in this study.
Furthermore, since the present light-front constituent quark model is rather simple,  
%and based on a naive constituent quark picture, 
the model cannot discuss the chiral limit of vanishing (effective) light-quark masses.

The effective interaction Lagrangian density for the 
quarks and pion in medium is given by, 
\begin{eqnarray}
  \mathcal{L}_\mathrm{eff} & = &  -ig^*\, 
  (\bar{q}\gamma^5\vec\tau q \cdot \vec\phi) \ \Lambda^* \ , 
\end{eqnarray}
where the coupling constant,~$g^*=m^*_{q}/f^*_{\pi}$, is obtained by the ''in-medium 
Goldberger-Treiman relation'' at the quark level, with $m^*_{q}$ and $f^*_{\pi}$  
being respectively the effective constituent quark mass 
and pion decay constant in medium, $\vec \phi$ 
the pion field~\cite{deMelo2002,Yabusaki2015}, and 
$\Lambda^*$ is the $\pi$-$q$-$\bar{q}$ vertex function in medium.
Hereafter, the in-medium quantities are indicated with the asterisk, $^*$.

%{\it Symmetric pion valence wave function:}~
The pion valence wave function used in this study   
is symmetric under the exchange of quark and antiquark momenta. 
This $\pi$-$q$-$\bar{q}$ vertex function, $\Lambda (k,P)$ in vacuum 
%with the arguments $k$ and $P$ stand for momenta, 
is the same as that used for studying the properties of pion~\cite{deMelo2002} 
and kaon~\cite{Yabusaki2015}. 
However, for the in-medium $\Lambda^*$, the arguments 
of the function are replaced by those of the in-medium~\cite{deMelo2014}:
\begin{equation}
\Lambda^*(k+V,P)=
\frac{C^*}{((k+V)^2-m^2_{R} + i\epsilon)}+
\frac{C^*}{((P-k-V)^2-m^2_{R}+ i\epsilon)},
\label{vertex}
\end{equation}
where $V^\mu = \delta^\mu_0 V^0$ is the vector potential felt 
by the light quarks in the pion immersed in medium, and can be eliminated by the 
variable change in the $k$-integration, $k^\mu + \delta^\mu_0 V^0 \to k^\mu$.
The normalization factor associated with $C^*$ is modified by the medium 
effects. 
%(See also below Eq.~(\ref{wf2}), and Ref.~\cite{deMelo2014} for details.) 
The regulator mass $m_R$ represents soft effects at short range of about the 1~GeV scale, 
and $m_R$ may also be influenced by in-medium effects. 
However, we employ $m_R^* = m_R$ in Eq.~(\ref{vertex}), 
since there exists no established way of estimating this effect
on the regulator mass. 
This can avoid introducing extra source of uncertainty.

The Bethe-Salpeter amplitude in medium, $\Psi_\pi^*$, with the 
vertex function in medium $\Lambda^*$ is given by,
\begin{eqnarray}
\Psi_\pi^*(k+V,P) & = & \frac{\rlap\slash{k}+\rlap\slash V+m_q^*}{(k+V)^2-m_q^{*2}+ i\epsilon}
\gamma^5 \Lambda^* (k+V,P) 
\frac{\rlap\slash{k}+\rlap\slash V-\rlap\slash{P}+m_q^*}{(k+V-P)^2-m_q^{*2}+ i\epsilon}.
\label{bsa}
\end{eqnarray}

By eliminating the instantaneous terms, or eliminating 
the terms with the matrix $\gamma^+$ in the numerators   
and $k^+$ and~$(P^+ - k^+)$ in the denominators 
with the light-front convention $a^{\pm} \equiv a^0 \pm a^3$, 
and integrating over the light-front energy~$k^-$,  
we obtain the in-medium pion valence wave function $\Phi_\pi^*$,     
\begin{eqnarray}
\hspace*{-2ex}
\Phi_\pi^*(k^+,\vec k_\perp; P^+,\vec P_\perp)=
\frac{P^+}{m^{*2}_\pi-M^2_0} 
\left[\frac{N^*}
{(1-x)(m^{*2}_{\pi}-{\cal M}^2(m_q^{*2}, m_R^2))} 
+\frac{N^*}{x(m^{*2}_{\pi}-{\cal M}^2(m^{2}_R, m_q^{*2}))} \right],
\label{wf2}
\end{eqnarray}
where, $N^*=C^* (m^*_{q}/f^*_{\pi}) (N_c)^{\frac{1}{2}}$ 
is the normalization factor with the number of colors $N_c$~\cite{deMelo2002,deMelo2014}, 
$x=k^+/P^+$ with~$0 \le x \le 1$,  
${\cal M}^2(m^2_a, m_b^2) \equiv  
\frac{\vec{k}^2_\perp+m_a^2}{x}
+\frac{(\vec{P}-\vec{k})^2_\perp+m^2_{b}}{1-x}-\vec{P}^2_\perp \ $,
the square of the mass $M^2_0$ is $M^2_0 ={\cal M}^2(m_q^{*2}, m_q^{*2})$, 
and $m_R$ is the regulator mass with the value $m^*_R = m_R = 600$ MeV~\cite{deMelo2002,deMelo2014}. 
Note that the model used in Refs.~\cite{deMelo1999,daSilva2012} 
does not have the second term in Eq.~(\ref{wf2}). 
This means that the pion valence wave function in Refs.~\cite{deMelo1999,daSilva2012} 
is not symmetric under the exchange of quark and antiquark momenta.

The pion transverse momentum probability density in medium,   
$P_\pi^*(k_\perp)$, in the pion rest frame $P^+=m_\pi^*$ is 
calculated by, 
\begin{eqnarray}
P_\pi^*(k_\perp)= \frac{1}{4\pi^3 m^*_\pi} \int_0^{2\pi} d\phi 
\int^{m_\pi^*}_0 \frac{d k^{+}M_0^{*2}}
{k^+(m_\pi^*-k^+)} |\Phi_\pi^*(k^+,\vec k_\perp;m^*_\pi,\vec 0)|^2, 
\label{prob1}
\end{eqnarray}
and the integration over $k_\perp$ for $P_\pi^*(k_\perp)$ leads to the in-medium probability 
of the valence component in the pion, $\eta^*$~\cite{deMelo2002,deMelo2014}:  
\begin{eqnarray}
\eta^*=\int^\infty_0 dk_\perp k_\perp P_\pi^*(k_\perp).
\label{eta}
\end{eqnarray}

\begin{figure}[tb]
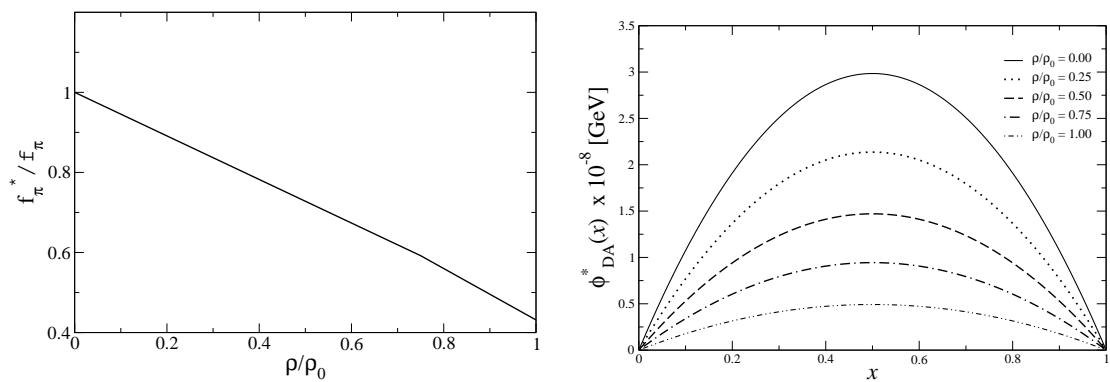

\mbox{
\epsfig{figure=fpistar2.eps,width=7cm}
\hspace{2ex}
\epsfig{figure=mdapiv1.eps,width=7cm}
} \par
\caption{Pion decay constant calculated in symmetric nuclear 
matter~(left panel) taken from Ref.~\cite{deMelo2014}, 
and pion valence distribution amplitudes (right panel),
taken from Ref.~\cite{piDA}.
}
\label{fpiDA}
%\vspace{3ex}
\end{figure}

The pion decay constant in medium, 
in terms of the pion valence component with 
$\Phi_\pi^*(k^+,\vec{k}_\perp;m^*_\pi,\vec{0})$, is calculate by~\cite{deMelo2002,deMelo2014}: 
\begin{eqnarray}
f^*_\pi = \frac{m_q^* (N_c)^{1/2}}{4\pi^3} 
\int \frac{d^{2} k_{\perp} d k^+ }{k^+(m_\pi^*-k^+)} 
\Phi_\pi^*(k^+,\vec{k}_\perp;m^*_\pi,\vec{0}).
\label{fpi}
\end{eqnarray}
The result is shown in Fig.~\ref{fpiDA} (left panel) as a function of nuclear density 
($\rho/\rho_0$). 
Note that, $f^*_\pi$ above is calculated with the plus-component of the light front 
axial-vector current (light-front time component). 
Thus the $f^*_\pi$ cannot be separated into the usual sense of the time and space 
components, where the corresponding two components of $f^*_\pi$ decouple 
with the presence of background matter (in nuclear medium), 
and the time component is shown to be model independent and directly associated 
with the observables, and also with the Gell-Mann-Okes-Renner 
(GMOR) relation~\cite{Kirchbach}. 
As nuclear density increases, the in-medium pion decay constant $f^*_\pi$ decreases,  
although the amount of the reduction may be larger than that calculated by the  
NJL model~\cite{Vogl,Lutz}.
However, in the treatment of the NJL model~\cite{Vogl,Lutz}, 
the decoupling of the time and space components of 
$f^*_\pi$ is not clear with the presence of background matter, 
and may not directly be compared with our result,  
as well as the empirical value extracted from 
the pionic-atom experiment~\cite{Kienle}.

In the following, we further discuss the in-medium ``quark condensate'' and 
the GMOR-like relation, just for an illustration. 
As we mentioned already, the present model is a light-front 
constituent quark model with the constituent quark mass of $m_q=220$~MeV in vacuum, 
and thus we cannot discuss the chiral limit (model limitation).
Keeping this point in mind, however, to get some idea, 
we simply write down the GMOR-like relation 
in vacuum and in medium, so that we try to compare the ``quark condensate'' value 
with that extracted experimentally~\cite{Kienle}:
\begin{eqnarray}
m^2_\pi f^2_\pi &=& - 2 m_q <\overline{q} q>,
\label{GMOR}\\
m^{*2}_\pi f^{*2}_\pi &=& - 2 m^*_q <\overline{q} q>^* \ .
\label{GMORm}
\end{eqnarray}
Then, the ratio of the in-medium to vacuum quark condensates in the present approach may be estimated as,
\begin{equation}
\frac{<\overline{q} q>^*}{<\overline{q} q>} = \frac{m_q}{m^*_q} \frac{m^{*2}_\pi f^{*2}_\pi}{m^2_\pi f^2_\pi}
\simeq \frac{m_q}{m^*_q} \frac{f^{*2}_\pi}{f^2_\pi} \ .
\label{qconr}
\end{equation}
At normal nuclear matter density, $\rho_0$ (0.15 fm$^{-3}$), the ratio gives 
$\simeq 0.52$~\cite{deMelo2014} 
(and also one can calculate by using the values listed in table~\ref{Tab:summary}, to be given next). 
This implies a larger reduction in ``quark condensate'' compared to
the value $0.67 \pm 0.06$ extracted in Ref.~\cite{Kienle} at a density 0.17 fm$^{-3}$ (their value for
the normal nuclear matter density). This feature may also be understood from the larger reduction in
$(f^*_\pi/f_\pi)^2$ in our approach compared with that obtained 
in Ref.~\cite{Kienle} (see also Ref.~\cite{deMelo2014} for more details). 
We repeat that the discussions above are not rigorous, but just for giving some intuition, 
since the vacuum structure in the light-front approach is usually considered as ``trivial'', and 
it is very difficult to study the quark condensate (or spontaneously broken chiral symmetry) 
in vacuum as well as in medium. This is a very interesting issue for the future elaboration 
within the light-front approach.

%%%
\begin{table}[t]
\begin{center}
\caption{
Properties of pion in medium, taken from Ref.~\cite{deMelo2014}, with   
$\rho_0=0.15$ fm$^{-3}$.
}
\label{Tab:summary}
%\vspace*{3mm}
\begin{tabular}{|c|c|c|c|c|}
\hline
$\rho/\rho_0$  & $m^*_q$~[MeV] & $f^*_{\pi}$~[MeV] & $<r^{*2}_{\pi}>^{1/2}$~[fm] & $\eta^*$ \\
\hline
~0.00  &  ~220    & ~93.1   & ~0.73   & ~0.782   \\
~0.25  &  ~179.9  & ~80.6   & ~0.84   & ~0.812    \\
~0.50  &  ~143.2  & ~68.0   & ~1.00   & ~0.843     \\
~0.75  &  ~109.8  & ~55.1   & ~1.26   & ~0.878      \\
~1.00  &  ~79.5   & ~40.2   & ~1.96   & ~0.930       \\ 
\hline
\end{tabular}
\end{center}
%\vspace{8ex}
\end{table}

%%%
Some properties of the pion in symmetric nuclear matter 
obtained in Ref.~\cite{deMelo2014}, are summarized in table~\ref{Tab:summary}.
The results listed in table~\ref{Tab:summary} are summarized as follows.
[See also Fig.~\ref{E/Amqstar} (right panel) for $m^*_q$, Fig.~\ref{ffradius} (right panel) 
for $<r^{*2}_\pi>^{1/2}$, and~Fig.~\ref{fpiDA} (left panel) for $f^*_\pi$.] 
As the nuclear density increases, $m_q^*$ and $f_\pi^*$ decrease, while  $<r_\pi^{*2}>^{1/2}$  
and the probability of valence component in the pion, $\eta^*$, increase.
This can be understood as follows.
The reduction in mass, $m_q^*$, makes it easier 
to excite the valence quark component in the pion state, and resulting to 
increase the valence component probability $\eta^*$ in the pion. 
Furthermore, the valence wave function spreads 
more in coordinate space by the decrease of $m_q^*$, 
and reduces the absolute value of the 
wave function at the origin ($f_\pi^* \propto |\Phi_\pi^*(\vec{r}=\vec{0})|$ 
reduction~\cite{Weisskopf}), namely, increases $<r_\pi^{*2}>^{1/2}$.

%{\it In-medium pion Distribution Amplitude:} 
Next we discuss the in-medium pion valence distribution amplitude.
Pion DA provides information 
on the nonperturbative regime of the bound state  
nature of pion due to the quark and antiquark at higher momentum transfer.
The pion valence wave function in vacuum is normalized 
by~\cite{Lepage:1982gd,Brodsky:1989pv} 
(aside from the factor $\sqrt{2}$ difference):
%%%
\begin{equation}
 \int_0^1 dx \int \frac{d^2 k_\perp}{16 \pi^3 }
 \Phi_{\pi}(x,\vec{k}_\perp)~=~ \frac{f_{\pi}}{2 \sqrt{6}}.
\label{normal}
\end{equation}
%%%
This is an important constraint on the normalization of the $q\bar{q}$ wave 
function~\cite{Lepage:1982gd,Brodsky:1989pv}, associated with a probability 
of finding a pure $q\bar{q}$ state in the pion state. 
According to this normalization, the in-medium pion valence wave function 
is normalized by replacing $f_\pi \to f_\pi^*$ in the above.
Since the pion decay constant in nuclear medium is modified, 
the pion valence wave function in nuclear medium is also modified 
via this normalization.

\begin{figure}[tb]
\begin{center}
\hspace*{-1.4cm}
\mbox{
\epsfig{figure=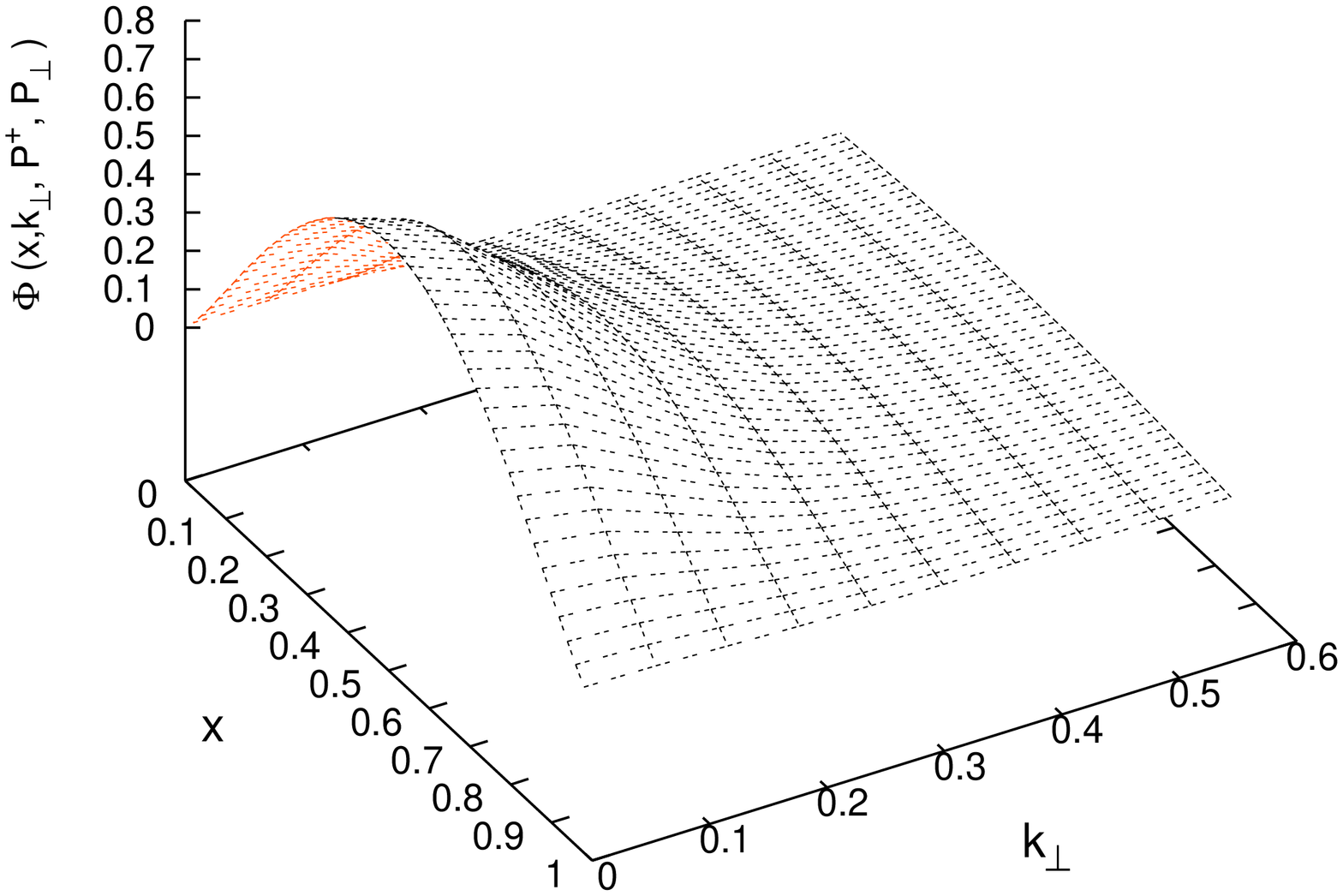,width=6.6cm,angle=-90}
\hspace{-2.3cm}
\epsfig{figure=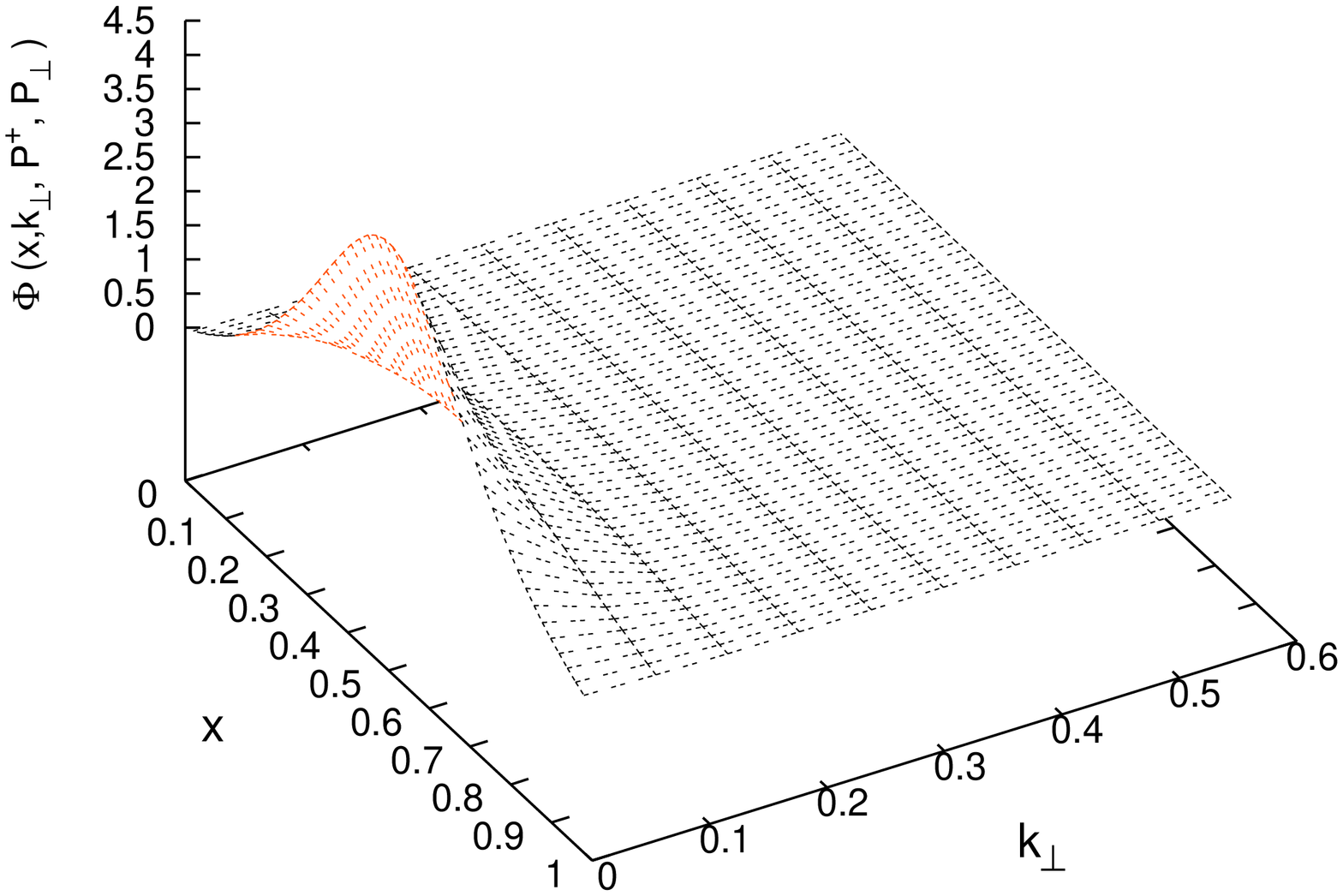,width=6.6cm,angle=-90}
} \par
\caption{Pion valence wave functions in vacuum ($\rho=0$) [left panel] 
and in medium ($\rho=\rho_0$) [right panel]  
v.s. $x$ and $k_\perp = |\vec{k}_\perp|$, where $P^+=m_\pi=m_\pi^*$ and 
$P_\perp=|\vec{P}_\perp|=0$.
The wave functions are given in the units, $10^{-8}\times$(GeV)$^{-1}$. 
Notice that the differences in the vertical axis scales 
for the left and right panels. (Taken from Ref.~\cite{piDA}.)
}
%\vspace{1.0cm}
\label{wf}
\end{center} 
\end{figure}

In order to examine more in detail as to how the change in $f_\pi^*$ impacts 
on the in-medium pion valence wave function, 
we show in Fig.~\ref{wf} the pion valence wave functions 
in vacuum (left panel) and $\rho=\rho_0$ (right panel).
%%%%%%%%%%%%%%%%%%
%%%%%%%%%%%%%%%%%%
%%%%%%%%%%%%%%%%%%
%%%%%%%%%%%%%%%%%%
%%%%%%%%%%%%%%%%%%
\begin{figure}[tb]
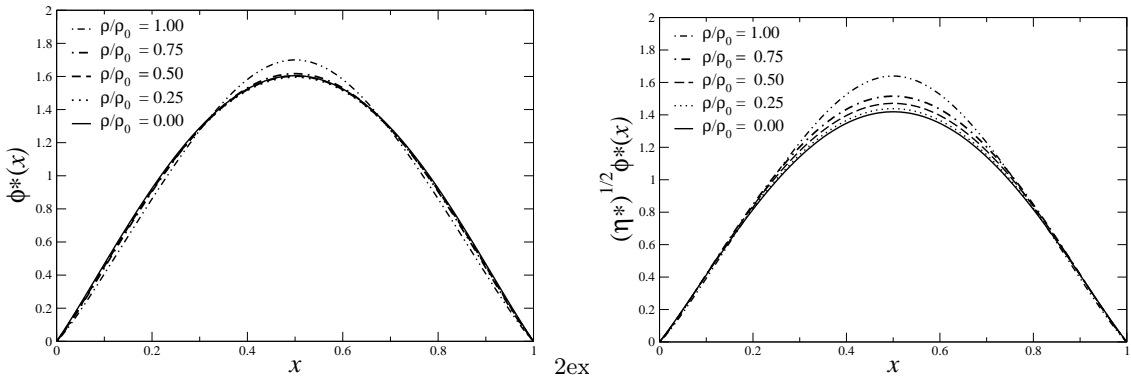

\begin{center}
\mbox{
\epsfig{figure=medpdafig1.eps,width=7.0cm}
\space{2ex}
\epsfig{figure=medpdafig3.eps,width=7cm}
} 
\end{center}
\par
\caption{Effective pion valence distribution amplitudes in vacuum and 
in medium, respectively multiplied by $\sqrt{\eta}$ and $\sqrt{\eta^*}$.
}
\label{EFFPDA}
\vspace{3ex}
\end{figure}
%%%%%%%%%%%%%%%%%%
One can notice that the in-medium pion valence wave function in momentum space 
has a sharper peak and localized in narrower regions both in $x$ and $k_\perp$   
than those in vacuum.
Of course, the total volume, the quantity integrated over $x$ and $\vec{k}_\perp$,  
is reduced in medium, corresponding to the reduced $f_\pi^*$.  
This fact is reflected in the wave function in coordinate space, that 
it becomes spread wider, and generally its hight is reduced.

The corresponding pion valence DA in medium, denoted by $\phi_{DA}^*(x)$  
(not normalized to unity), is calculated as 
\begin{equation}
 \phi_{DA}^*(x)=
 \int \frac{d^2k_{\perp}}{16\pi^3}\Phi_{\pi}^*(x,\vec{k}_\perp).
 \label{Eq:PDA}
\end{equation}
Note that, Eq.~(\ref{Eq:PDA}) holds also for the other pseudoscalar 
mesons $M_{ps}$ such as kaon and D-meson, by replacing 
$\Phi_{\pi}^*(x,\vec{k}_\perp) \to \Phi_{M_{ps}}^*(x,\vec{k}_\perp)$ in the above.

We show in Fig.~\ref{fpiDA} (right panel) the obtained pion DAs, $\phi_{DA}^*(x)$,  
in vacuum ($\rho/\rho_0=0$) and in medium for several nuclear densities. 
The significant reduction of the in-medium 
pion valence DA is obvious, that reflects the reduction of $f_\pi^*$ (left panel).

Next, we study pion valence DA normalized to unity, 
or normalized pion valence DAs in vacuum and in medium. 
By this, we can study the change in shape due to the medium effects.
We show in Fig.~\ref{EFFPDA} (left panel)  
the normalized pion valence DAs,  $\phi^*(x)$,  
both in vacuum ($\rho/\rho_0=0$) and in medium.
The in-medium change in shape is moderate  
when the nuclear matter densities are small, 
but it becomes evident when the density becomes $\rho_0$.

Furthermore, it may be useful to define {\it effective pion valence DA} 
using the valence probability in vacuum $\eta$ 
and in medium $\eta^*$. (See Eq.~(\ref{eta}) and table~\ref{Tab:summary}.) 
The pion states in vacuum, $|\pi>$, and in medium, $|\pi>^*$, can respectively 
be written as,
\begin{eqnarray}
|\pi> &=& \sqrt{\eta}|q\bar{q}> + a|q\bar{q}q\bar{q}> + b|q\bar{q}g> + \cdots,
\label{pistate0}\\
|\pi>^* &=& \sqrt{\eta^*}|q\bar{q}>^* + c|q\bar{q}q\bar{q}>^* + d|q\bar{q}g>^* + \cdots,
\label{pistate}
\end{eqnarray}
where $a, b, c$ and $d$  are constants, and $g$ denotes a gluon, 
and $+ \cdots$ stands for the higher Fock components in the pion states.
The quantity $\eta^*$ in table~\ref{Tab:summary} indicates that the valence $q\bar{q}$
component in the pion state increases in medium as nuclear density increases.
The effective pion valence DAs, $\sqrt{\eta^*}\phi(x)^*$, in vacuum ($\rho/\rho_0=0$) and in medium,  
are shown in Fig.~\ref{EFFPDA}. They may respectively 
correspond to the first terms of Eqs.~(\ref{pistate0}) and~(\ref{pistate}).

Since $\eta^*/\eta$ is enhanced in medium, effective pion valence DA   
in medium is also enhanced, on the top of the corresponding 
medium-(shape)modified normalized pion valence DA. 
The obvious enhancement of the effective pion valence DA in medium can be seen around $x=0.5$. 
This quantity may be useful when one studies some reactions in medium (in a nucleus)  
involving a pion, based on a constituent quark picture of pion.

\section{Summary}
\label{summary}

We have studied the impact of in-medium effects on the pion valence 
distribution amplitudes using a light-front constituent quark model, 
combined with the in-medium input  
calculated by the quark-meson coupling model. 
The in-medium constituent light-quark properties inside the pion are consistently 
constrained by the saturation properties of symmetric nuclear matter. 

The in-medium pion mass is assumed to be the same as that in vacuum, based on the 
extracted information from the pionic-atom experiment, and some theoretical studies. 
This information extracted is valid up to around the normal nuclear matter density. 
Thus, the results obtained in this study, combined with the light-front constituent 
quark model, are valid up to around the normal nuclear matter density, 
but cannot discuss reliably the chiral limit, the vanishing limit of 
the (effective constituent) light-quark masses. We need to rely on more sophisticated models 
of pion to be able to discuss the chiral limit in medium, as well as in vacuum.

Due to the reduction in the pion decay constant 
in medium, the pion distribution amplitude in medium normalized with  
the pion decay constant, is substantially reduced at relatively higher nuclear densities.
Because the valence component probability in medium increases as nuclear 
density increases, we have defined an effective pion distribution amplitude 
normalized to the square root of the valence probability in the pion state. 
This may give some information for the effectiveness 
of the valence quark picture of pion in nuclear medium. 
Within the present light-front constituent quark model approach, 
the effectiveness of the valence quark picture of the pion in medium, becomes 
more enhanced as nuclear density increases, due to the increase of the 
valence component in the pion state.

Although the present study is based on a simple, light-front constituent 
quark model, this is a first step to understand the impact of the 
medium effects on the internal structure of the pion immersed in nuclear medium.
%In the future, we plan to make similar studies for kaon, D-meson, and $\rho$-meson 
%in nuclear medium.

\begin{acknowledgements}
The authors would like to thank the organizers of Light Cone 2016 for 
invitation, and warm hospitality, that motivated them and 
made them enjoy during the workshop.
\end{acknowledgements}

% BibTeX users please use
%\bibliographystyle{spbasic}


\end{document}